\documentclass[namedreferences]{solarphysics}
\usepackage[hyperref,optionalrh,solaromanenum]{spr-sola-addons} 
\usepackage{graphicx}                    
\usepackage{color}                       
\usepackage{breakurl}                         

\begin{document}

\begin{article}

\begin{opening}

\title{Common origin of quasi-periodic pulsations in microwave and decimetric solar radio bursts}

%
\author[addressref={aff3},corref,email={lkk@iszf.irk.ru}]{\inits{L.K.~}\fnm{Larisa~K.}~\lnm{Kashapova}\orcid{0000-0003-2074-5213}}
 
 \author[addressref={aff2,aff3},corref,email={d.kolotkov.1@warwick.ac.uk}]{\inits{D.Y.~}\fnm{Dmitrii~Y.~} \lnm{Kolotkov}\orcid{0000-0002-0687-6172}}
\author[addressref={aff1,aff3},corref,email={elenku@bk.ru}]{\inits{E.G.}\fnm{Elena~G.~}\lnm{Kupriyanova}\orcid{0000-0001-9664-0552}}
\author[addressref={aff3},corref,email={kudryavtsevau@iszf.irk.ru}]{\inits{A.V.}\fnm{Anastasiia~V.}~\lnm{Kudriavtseva}\orcid{0000-0003-3248-5498}}
\author[addressref={aff4},corref,email={tanchm@nao.cas.cn}]{\inits{C.M.}\fnm{Chengming~}\lnm{Tan}\orcid{0000-0001-5359-6897}}
\author[addressref={aff5,aff6},corref,email={hamish.reid@ucl.ac.uk}]{\inits{H.A.S.}\fnm{Hamish~A.~S.~}\lnm{Reid}\orcid{}}
%

\address[id=aff3]{Institute of Solar-Terrestrial Physics, SB Russian Academy of Sciences, 664033 Irkutsk, Russia}
\address[id={aff2}]{Centre for Fusion, Space and Astrophysics, Physics Department, University of Warwick, Coventry CV4 7AL, United Kingdom}

\address[id={aff1}]{Central Astronomical Observatory at Pulkovo, Russian Academy of Sciences, 196140 Saint Petersburg, Russia}

\address[id={aff4}]{CAS Key Laboratory of Solar Activity, National Astronomical Observatories of Chinese Academy of Sciences, Datun Road A20, Chaoyang District, Beijing 100101, China}
\address[id={aff5}]{Department of Space and Climate Physics, University College, London, RH5 6NT, UK} 
\address[id={aff6}]{School of Physics and Astronomy, University of Glasgow, Glasgow G12 8QQ, UK}

\begin{abstract}

We analyse quasi-periodic pulsations (QPP) detected in the microwave and decimeter radio emission of the SOL2017-09-05T07:04 solar flare, using simultaneous observations by the Siberian Radioheliograph 48 (SRH-48, 4--8 GHz) and Mingantu Spectral Radioheliograph  (MUSER-I, 0.4--2 GHz).  The microwave emission was broadband with a typical gyrosynchrotron spectrum, while a quasi-periodic enhancement of the decimetric emission appeared in a narrow spectral band (500--700 MHz), consistent with the coherent plasma emission mechanism.
The periodicity that we found in microwaves is about 30~s, coming from a compact loop-like source with a typical height of about 31\,Mm. The decimetric emission demonstrated a periodicity about 6~s.
We suggested a qualitative scenario linking the QPPs observed in both incoherent and coherent spectral bands and their generation mechanisms.
The properties of the QPPs found in the microwave signal are typical for perturbations of the flare loop by the standing sausage mode of a fast magnetohydrodynamic (MHD) wave. Our analysis indicated that this sausage-oscillating flare loop was the primary source of oscillations in the discussed event.
The suggested scenario is that a fundamental sausage harmonic is the dominant cause for the observed QPPs in the microwave emission.  The initiation of oscillations in the decimetric emission is caused by the third sausage harmonic \textit{via} periodic and nonlinear triggering of the acceleration processes in the current sheets, formed at the interface between the sausage-oscillating flare loop and the external coronal loop that extended to higher altitudes.
Our results demonstrate the possible role of MHD wave processes in the release and transport of energy during solar flares, linking coherent and incoherent radio emission mechanisms.

\end{abstract}

%
\end{opening}





%
 \section{Introduction}
 \label{s:Introduction} 
 
 Flares are the most complicated and powerful phenomena in the solar atmosphere. 
Flare energy release occurs in the corona but the effects can extend down to the photosphere, and out to the upper corona and interplanetary space.
 {Such a} wide range of heights 
 {with significantly different}
 plasma conditions results in electromagnetic emission ranging from visible and UV to radio, {X- and $\gamma$-ray spectral bands} \citep[see, for example,][]{Fl2011SSRv..159...19F}. Despite different energetic{s} and generation mechanisms, {manifestations of solar flares} in different spectral {bands} are undoubtedly related. However, the {specific physical} mechanisms of these relations{hips} are still unclear {and remain 
 {an}
 open and actively debated research topic}.
 {One of the most 
 {widely} 
 known examples of such correlations between the multi-band flare emissions is the empirically established Neupert effect: {The connection between the} non-thermal X-ray and microwave emission time profiles, and the derivative of 
 the thermal
 soft X-ray time profiles \citep{Neupert1968ApJ...153L..59N}.}
 Another 
 correlation
 example is the solar radio type III bursts in {the} meter radio {band} 
 and decimetric spike
 bursts. Both radio bursts are associated with electrons accelerated during flares or flare-like events. There are two possible origins of {the} accelerated electrons that produce this radio emission. Electrons can be accelerated {by the primary flare energy release} and 
 {travel along}
 open magnetic field lines where they can escape the solar atmosphere. Alternatively, accelerated electrons can be produced in {smaller-scale, local} current sheets {formed by interacting} magnetic structures {in the same active region}. {For example,}  \citet[][]{Benz2009A&A...499L..33B} {and} \citet[][]{Benz2011SoPh..273..363B} {demonstrated} that the position of decimetric {emission} sources did not agree with the model of electrons {trapped in a flare loop}. {The} generation of accelerat{ed} electrons by current sheets {was suggested as a} more realistic {scenario}. {On the other hand,} it {raises} the question {of} the {connection between the decimetric emission produced by these electrons and the flare emission in the microwave and X-ray bands.}
 
The microwave (MW) {radiation from} solar flare{s} is {usually associated with} gyrosynchrotron emission from relativistic {(or nearly relativistic)} electrons. {It} is {an} incoherent emission at frequencies usually above 1\,GHz, controlled by {the} accelerated electron properties and the strength of the magnetic field in the emission source \citep[see][]{Dulk1985ARA}. {The} MW gyrosynchrotron emission {is thought to be} more sensitive {to} accelerated electrons than hard X-rays. At lower frequencies, the emission is usually associated with a two-stages coherent plasma mechanism. 
{Electron beams with energies of tens of keV become unstable to Langmuir wave generation.  Radio waves are then produced via the nonlinear transformation of these Langmuir waves into electromagnetic emission at frequencies around the local plasma frequency and its second harmonic}
\citep[see][as a review]{2014RAA....14..773R}.

{The existence of }temporal correlations between coherent and incoherent flare radio emission{s has been known} since earlier studies \citep[\textit{e.g.}][]{1959AuJPh..12..399N}, where type III bursts were found to be generally accompanied by {enhanced} microwave {and hard X-ray} {flux}. {Assuming that} both types of emission {are induced} by the same population of accelerated electrons, analysis of the time delays between {them} allows{, in particular, for} detecting the direction of the electron beam propagation and estimating its speed. 
{For example, the temporal connection between type III spikes, hard X-ray spikes, and microwaves were studied by \citet{1984SoPh...90..383D, 1995ApJ...455..347A, 2014A&A...567A..85R, 2016SoPh..291.3427K}.}

 
An important {and perhaps intrinsic feature of solar flares} is quasi-periodic pulsations (QPPs), frequently seen in the time profiles of flare emission{ across the entire electromagnetic spectrum}. Statistical studies have shown that QPPs are present in both non-thermal \citep{2010SoPh..267..329K} and thermal \citep{2015SoPh..290.3625S} emissions of more than 80\% of the {analysed} solar flares {\citep[see also][for a more recent statistics of QPP events]{2020ApJ...895...50H}}. QPPs are {usually} considered as manifestations of magnetohydrodynamic (MHD) {wave} processes in various magnetic {plasma} configurations {in or nearby the flaring site}. 
Simultaneous analysis of QPP {properties} in different spectral bands, {such as} their timescales, amplitudes, and phases, {opens up new and promising opportunities for studying} the processes of energy release and energy transport {throughout the solar atmosphere} \citep[see][for {the most} recent reviews]{2020STP.....6a...3K, 2021SSRv..217...66Z}.

{Signatures of QPPs are regularly observed in coherent and incoherent radio emissions. For example,
\citet[][]{2009ApJ...697L.108M} and \citet[][]{2011SoPh..273..393M} detected QPP signals with periods about 71\,s and 81\,s in the type IV bursts at 1.1--4.5 GHz, which were associated with the modulation of the gyrosynchrotron emission by a packet of fast magnetoacoustic waves propagating in an open plasma structure. The magnetic field in the vicinity of the observed radio emission sources was estimated around 11\,G and 48\,G. \citet[][]{2017ApJ...844..149K} observed similar $\sim$70-s and $\sim$140-s quasi-periodicity in the type IV and type III radio bursts at 245--610\,MHz and in soft X-rays. Moreover, these radio-band QPPs were found to occur simultaneously with a coronal fast magnetoacoustic wave train, directly observed in Extreme Ultraviolet (EUV) images. Likewise, signatures of standing fast magnetoacoustic waves of a sausage symmetry with periods between $\sim$1\,s and $\sim$15\,s were found as QPP signals in coherent \citep[\textit{e.g.}][]{2018ApJ...859..154N} and incoherent \citep[\textit{e.g.}][]{2015A&A...574A..53K} radio emissions from weak (B-class) and strong (X-class) solar flares, respectively.}
{For more details on the dynamics of fast sausage waves in the corona and their role in modulation of the flare emission in the radio band, see the most recent review by \citet{2020SSRv..216..136L}.}

Higher {up} in the corona, at radio frequencies 42--83\,MHz, \citet[][]{2016A&A...594A..96G} detected quasi-periodic {\lq\lq sparks\rq\rq\ in} the type II burst, {whose} period in {the} radio band (1.78~min) matches the period of {a quasi-periodic fast-propagating} wave train observed in images of the lower {corona} in {the} EUV band. Moreover, 
{the speed of the emission location of the observed radio sparks,} estimated from the dynamic {radio} spectrum, {was found to be comparable} to the speed of the coronal mass ejection{, inferred} from {the} EUV data.
{\citet{2016SoPh..291.3427K} found QPPs present in MWs, hard X-rays and type III bursts with a common period of 40--50~s. Their findings indicated that the detected periodicity was probably associated with periodic dynamics in the injection of non-thermal electrons, which can be produced by periodic modulation of magnetic reconnection.}
Another {example of the modulation of the low-frequency radio emission by} the {fast MHD} wave was found by \citet{2018ApJ...861...33K} as quasi-periodic drifting fine structures in a type III radio burst at 33--40~MHz, using data {from} the LOw Frequency ARray (LOFAR). {The observed wave period of $\sim$3\,s and apparent wave propagation speed were used to probe the coronal magnetic field at $\sim$1.7 solar radii and compare the energy flux carried by the wave to local radiative losses.}

In this work, we aim to reveal the role of MHD wave processes in the energy release and transport in solar flares. {Our study develops the idea of \citet{Benz2011SoPh..273..363B} that the electrons may get accelerated not  in the site of primary energy releases  but also due to local interactions of the magnetic plasma structures residing in the active region.
For this purpose, we applied the analysis of QPP signals simultaneously detected in the MW and decimetric emission from a C-class solar flare SOL2017-09-05T07:04  as a tool for studying possible correlations and interplay between the mechanisms producing both types of emission.} 

The paper is organised as follows. {The description of the spatially, temporally, and spectrally resolved observational data and their analysis} are {given} in Section~\ref{s:Observations}. The {detected} properties of the {MW and decimetric} QPP {events} are described in Section~\ref{s:qpp}. The discussion of the obtained {observational} results {and future perspectives are outlined in} Section~\ref{s:DiscussionC}.


\section{Observations: data and analysis }
\label{s:Observations}

The {analysed} event occurred in the active region (AR) 12673, which produce{d a} series of powerful flares in September 2017. The event started at 
{around} 
07:04~UT 05 September 2017, {approximately 25\,min after} the M3.6 GOES class solar flare, and had a GOES class of C6.9. 

\subsection{Siberian Radioheliograph-48}
\label{s:Data_SRH}

Our study of the microwave flare emission is based on the data from the Siberian Radioheliograph-48 \citep[SRH-48,][]{Lesovoi2014RAA,Lesovoi2017STP}. In 2016, the prototype of a T-shaped 48-antenna array (SRH-48) started observations within the 4--8~GHz frequency band. The instrument uses an imaging approach based on the Fourier synthesis~\citep{Lesovoi2017STP}. Both circularly-polarised components ($R$ and $L$) are measured. SRH-48 observed at five frequencies~--- 4.5, 5.2, 6.0, 6.8, and 7.5~GHz from July 2016 to May 2018. {In the current study, we used only intensity ($R+L$) values.} The bandwidth of the frequencies is 10~MHz. 
The switch time between the frequencies was about 2~s, while the accumulation time at each frequency was 0.28~s. The best spatial resolution was 70'', achieved at 8~GHz. The image processing was done by the original software developed and tested to produce raw solar images, clean and calibrate them in brightness temperature. The method of image calibration in brightness temperatures is described by \cite{Kachnov2013PASJ}.
We note that the obtained brightness temperatures of compact sources can be considerably lower than the original temperatures if the source is less than the beam pattern of SRH-48.

For a study of temporal evolution, we used so-called correlation plots. They represent a proxy of radio ﬂux and display a temporal variation of the sum of cross-correlations of all antenna pairs. This type of data is very convenient for solar activity monitoring and are characterised by very high sensitivity to variations both in the brightness of compact sources and in their structure. Real-time correlation plots and quick-look images produced by SRH at ﬁve frequencies are accessible online at \url{http://badary.iszf.irk.ru/srhCorrPlot.php}.
Methods for calculating the correlation plots and their relation with characteristics of solar emission are presented in \cite{LesovoiKobets2017STP}. The authors showed that the sensitivity of SRH-48 to the emission of compact sources was up to 10$^{-2}$~s.f.u., and that 1\% increase in the correlation corresponds to an increment of about 5--10~s.f.u. within {the 4--8~GHz band}.

\subsection{Mingantu Spectral Radioheliograph}
\label{s:Data_MUSER}
The Mingantu Spectral Radioheliograph (MUSER) observes the Sun with high temporal, spatial and spectral resolutions in a wide range of frequencies \citep{Yan2016IAUS..320..427Y,Yan2021FrASS...8...20Y}. It consists of two instruments: MUSER-I and MUSER-II.
The spectral band of MUSER-I is 0.4--2~GHz. It has a spatial resolution of 51.6--10.3'' and 25~ms temporal resolution. MUSER-II observes within 2--15~GHz with 10.3--1.3'' spatial resolution and  206.25 ms temporal resolution. The data are processed using the original software, including calibration and image processing \citep{Yan2002ScChA..45...89Y,Wang2013PASJ...65S..18W,Chengming2015ApJ...808...61T}. In total, MUSER has obtained more than 390~TB of observational data during 2014--2019. There are 85~solar radio burst events, and among them, more than 60~radio burst events contain fine structure. In the current study, we used the dynamic spectrum obtained by  MUSER-I with high spectral and temporal resolution.

\subsection{Microwave spectral data}
\label{s:Data_sp}
Analysing oscillations in solar flare emission, we must be sure that the detected periods are not instrumental. {To ensure their solar origin, we analysed the same spectral band using independent instruments.  The first instrument is the 
{BroadBand Microwave Spectropolarimeters (BBMS),} 
observing within 3.8--8.2~GHz \citep{SSRS2011CEAB,SSRS2015SoPh}. {BBMS covers this frequency range using 16 channels with a cadence of 0.6~s}.  
The second} instrument providing observations of the solar radio flux at similar frequencies is the Radio Solar Telescope Network (RSTN).  We used data from the Learmonth station obtained within 245~MHz and 15.4~GHz with a 1-s cadence \citep[see, for example,][]{Kennewell:1983aa}.

\subsection{EUV  and X-ray observations}
\label{s:Data_EUV}

Due to an eclipse of the Solar Dynamics Observatory, the analysed solar flare SOL2017-09-05T07:04 was only observed in the EUV band by the Sun Watcher with Active Pixel System detector and Image Processing (SWAP) telescope onboard the ESA PROBA2 \citep{SWAP2013SoPh..286...43S,SWAP2013SoPh..286...67H}. We used SWAP images obtained in the 174~{\AA} band with a spatial resolution of 3.16'' \textit{per} pixel.
 
In the X-ray band, the event was observed by GOES only.
We used time profiles of the emission within the 1--8~{\AA} band that is sensitive to thermal processes, {and the derivative of this time profile} for studying the behaviour of accelerated electrons. 

\subsection{Event description and analysis of spectral and spatial  features of the flare sources}

%
\begin{figure} 
 \centerline{ \includegraphics[width=\textwidth,clip=]{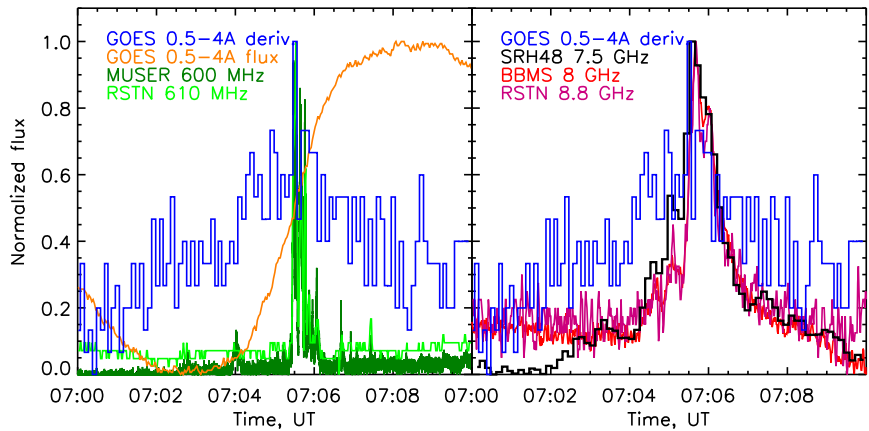}  }
 \caption{Comparison of microwave, decimeter radio and X-ray emission time profiles. The left panel: The evolution of emission at 600~MHz by MUSER and RSTN observations \textit{vs} soft X-ray time profile within 0.5--4~{\AA} band by GOES and its derivative. The right  panel: The evolution of the MW flux observed by SRH-48, BBMS, and RSTN at about 8~GHz \textit{vs} the derivative of the soft X-ray time profile within 0.5--4~{\AA} by GOES.}
 \label{fig:GOES}
\end{figure}

The comparison of the radio observations to the X-ray data is shown in  Figure~\ref{fig:GOES}. We 
{present}
the time profiles in the decimeter radio and MW bands from several independent instruments; the original SRH-48 correlation plot, BBMS, MUSER, and RSTN.
The cross-instrumental analysis shows a close correlation between the time profiles obtained by different remote instruments at similar frequencies. This fact allows us to conclude that the event and its fine structure had a solar 
{origin and was not related to instrumental or terrestrial ionospheric effects.}
As we can see in Figure~\ref{fig:GOES}, both bursts occurred during the rise phase of the flare.

We analysed the derivative of the soft X-ray time profile (SXR) in the 0.5--4~{\AA} band for the role of accelerated electrons in the generation of the analysed emissions \citep[Neupert effect,][]{Neupert1968ApJ...153L..59N}. One can note a close agreement of both MW and decimeter radio bursts with the derivative of the SXR time profile. Moreover, we see that the 600~MHz burst 
{was shorter in duration and}
occurred at the maximum of the time profile of the SXR derivative. These facts indicate the presence of accelerated electrons and their possible contribution to the generation of radio emissions.
Comparison of the decaying phase of the MW  and the SXR derivative time profiles demonstrates the absence of the trap effect {\cite[see, for example,][]{Melnikov1994R&QE}. Thus, we can exclude the magnetic trap from the flare topology and the model \lq\lq trap plus precipitation\rq\rq\ from interpretation \cite[see,][]{Silva2000ApJ}.}

\begin{figure} 
 \centerline{
 \includegraphics[width=0.49\textwidth,clip=]{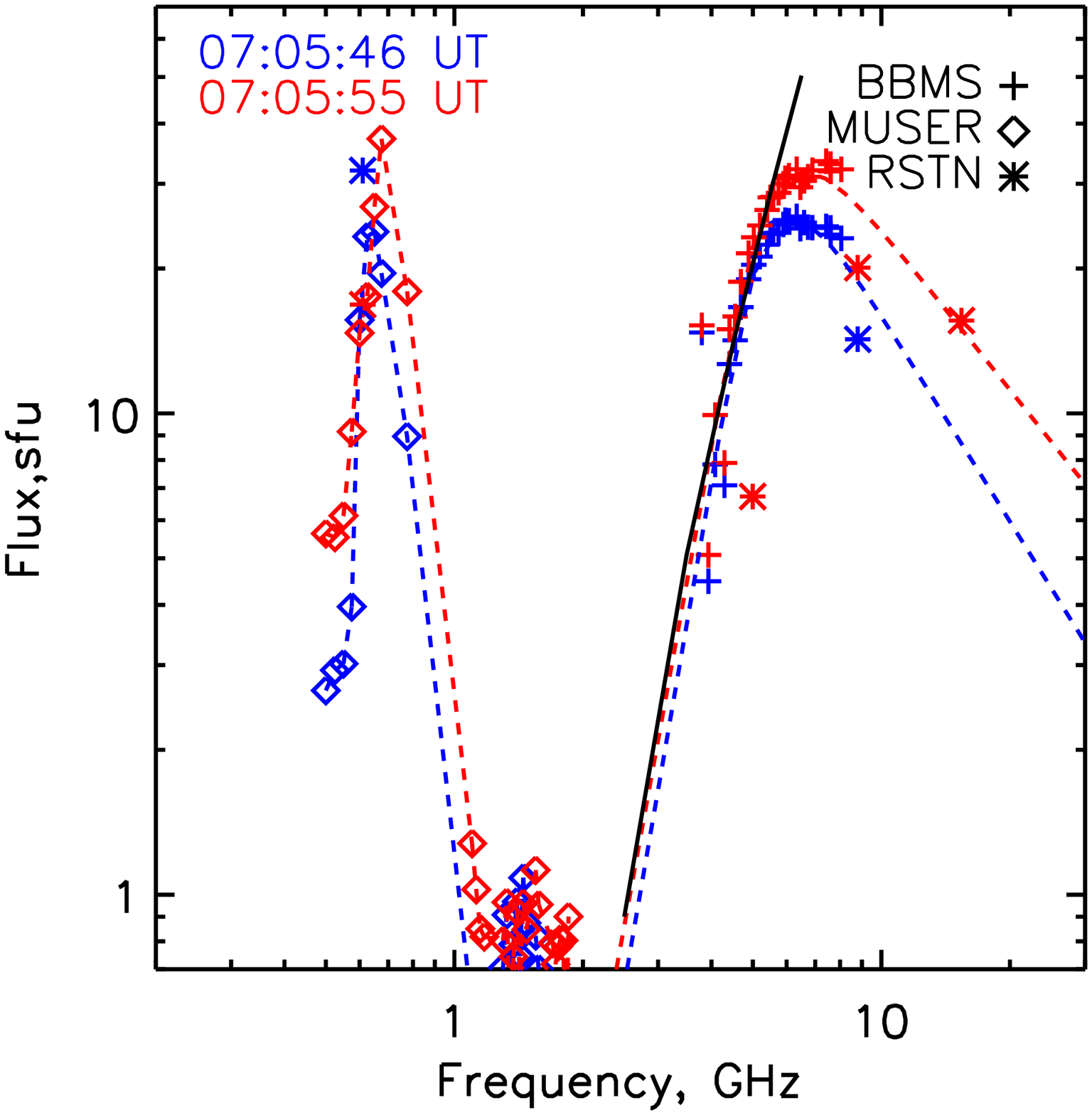}
 \includegraphics[width=0.49\textwidth,clip=]{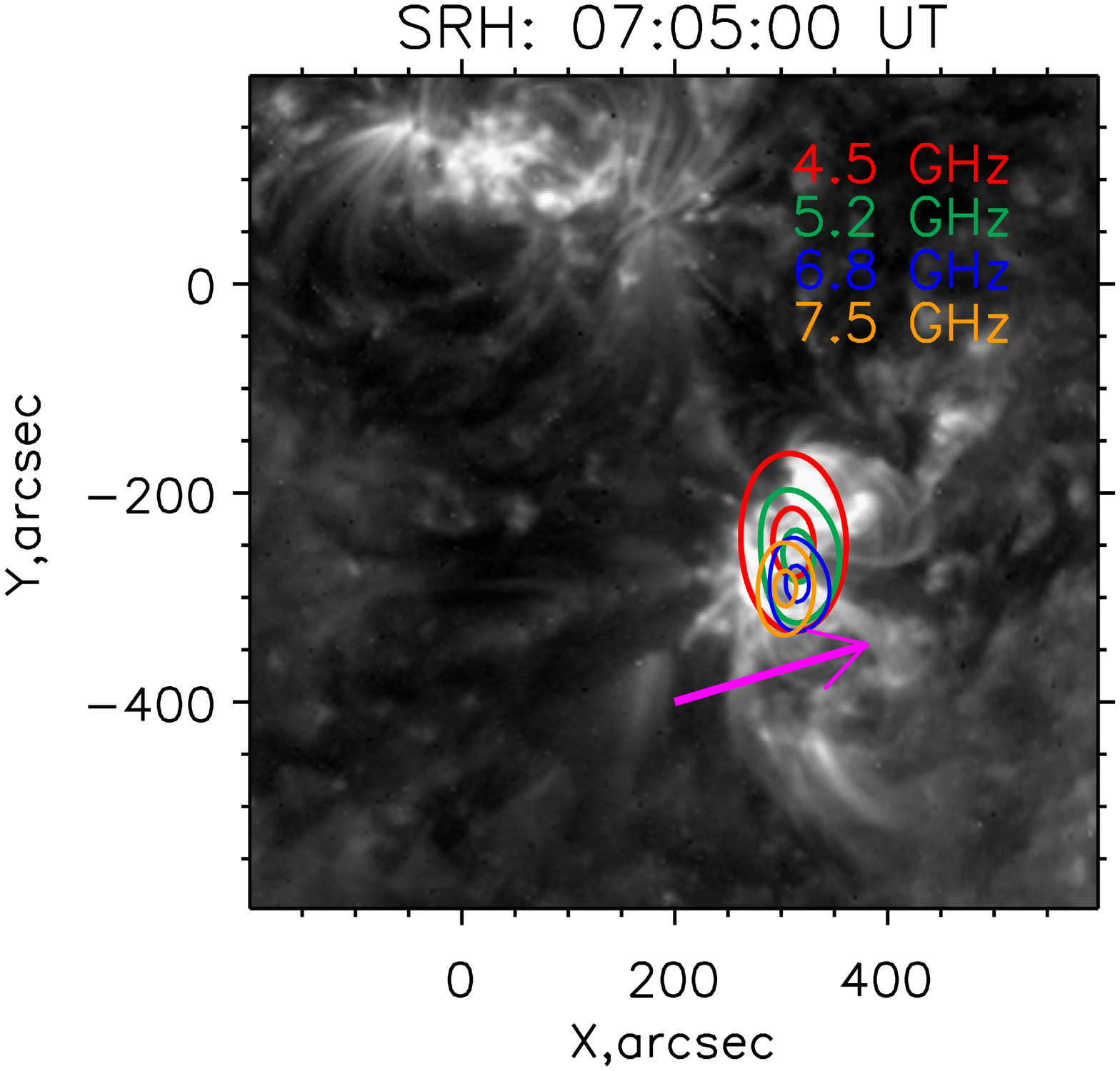}
  }
  \caption{The left panel: Radio spectra with {two maxima around 600 MHz and 6.6 GHz}, obtained from the data by BBMS, MUSER, and RSTN.  The blue colour shows flux at 07:05:46~UT (peak of the 600~MHz flux by RSTN). The red colour  marks data at about 07:05:55~UT (peak of the MW flux). The black line is $f^{2.5}$ function. The right panel: The SWAP 174~\AA\ image with the overlaid MW sources obtained by SRH-48. The sources are defined as the difference between the image at 07:05~UT and the image at 07:00~UT when flare activity in MW was absent. }
 \label{fig:EUV}
\end{figure}

We constructed the radio spectra from 100~MHz to 20~GHz using data from MUSER, BBMS, and RSTN for two instants in time. The first moment was chosen at the maximum of the 600~MHz time profile by RSTN. The left panel of  Figure~\ref{fig:EUV} shows this spectrum in blue.
{The second moment was chosen at the maximum of the MW burst, with the spectrum shown in red in Figure~\ref{fig:EUV}.}
Both MW spectra have a typical gyrosynchrotron form. The spectrum nearby 600~MHz is in turn very narrow-band and shows the form characteristic for the plasma emission mechanism. 

We obtained the peak frequency of the MW spectrum (the frequency of the maximum flux value) and the high-frequency photon spectral index using the technique by \citet{GELU2004ApJ...605..528N}. The peak frequency is found to be about 6.6~GHz for both moments of time considered. Thus, the SRH data at 6.8~GHz and 7.5~GHz corresponds to the optically thin part of the spectrum and indicates the segments of the flare loop near the footpoints. The MW sources at 4.5~GHz and 5.2~GHz in turn correspond to the optically thick emission from the entire flare loop. The photon spectral indices at high MW frequencies are found to be about 2.9 and 2.5 for the first and second moments, respectively. {These values are estimates for the non-thermal electron spectral  power-law index}. The spectral index at low MW frequencies is about 2.5 (see Figure~\ref{fig:EUV}), indicating an absence of the Razin effect and a normal (not anomalously high) density within the emitting loop \citep[see, for example,][]{Dulk1985ARA}.

AR 12673 is well-known for strong magnetic fields and well-developed magnetic structures, which resulted in a series of powerful solar flares \cite[see \textit{e.g.}][]{Wang2018RNAAS...2....8W, Kudriavtseva2021STP.....7a...3K}. To exclude the contribution of cyclotron emission from the sunspots and other permanent MW sources, we subtracted the image obtained at 07:00~UT (\textit{i.e.} before the analysed flare) from the images obtained during the MW burst. 
The MW sources which appeared at 07:05~UT are overlaid as {colour contours on a background} 174~{\AA} image by SWAP (see the right panel in Figure~\ref{fig:EUV}). The 6.8~GHz and 7.5~GHz images indicate a compact source coming from the vicinity of a flare loop footpoint, while the images at lower frequencies 4.5~GHz and 5.2~GHz demonstrate the entire flare loop structure. The distance between the centroids of 4.5~GHz and 7.5~GHz sources is about 43'' or 31~Mm (assuming 1'' is about 725~km). {Assuming the flare loop geometry is a half-circle, we can consider this distance between the centroids as the loop height.}
This value agrees with the size of an average flare loop usually seen in X-rays. Moreover, the previous studies of decimetric pulsation positions relative to coronal X-ray source {position} revealed that they are spatially separated by hundreds of arcseconds. This fact indicates that the radio burst emission was not generated by the electrons trapped in the flare loop, but likely came from a current sheet situated well above the X-ray loop top \citep{Benz2009A&A...499L..33B, Benz2011SoPh..273..363B}. 

We note that the sources at 6.8~GHz and 7.5~GHz are seen in Figure~\ref{fig:EUV} to situate near the fan-like structure clearly visible in the 174~{\AA} image by SWAP (marked by the magenta arrow in Figure~\ref{fig:EUV}, right panel). This structure was apparently long-lived and also visible in 171~{\AA} images of SDO/AIA before and after the time interval analysed in this work.

\begin{figure} 
 \centerline{
 \includegraphics[width=0.5\textwidth,clip=]{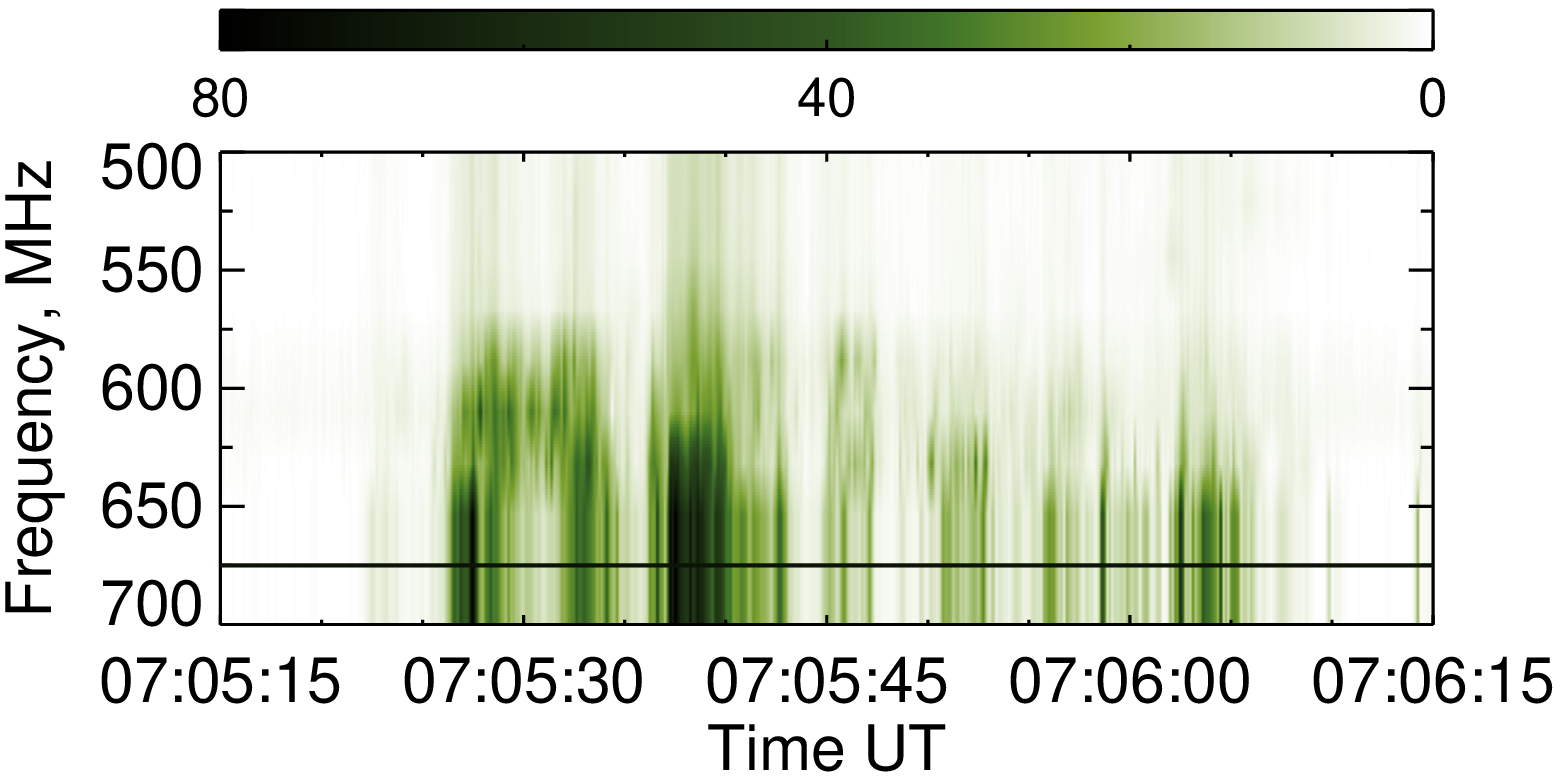}
 \includegraphics[width=0.5\textwidth,clip=]{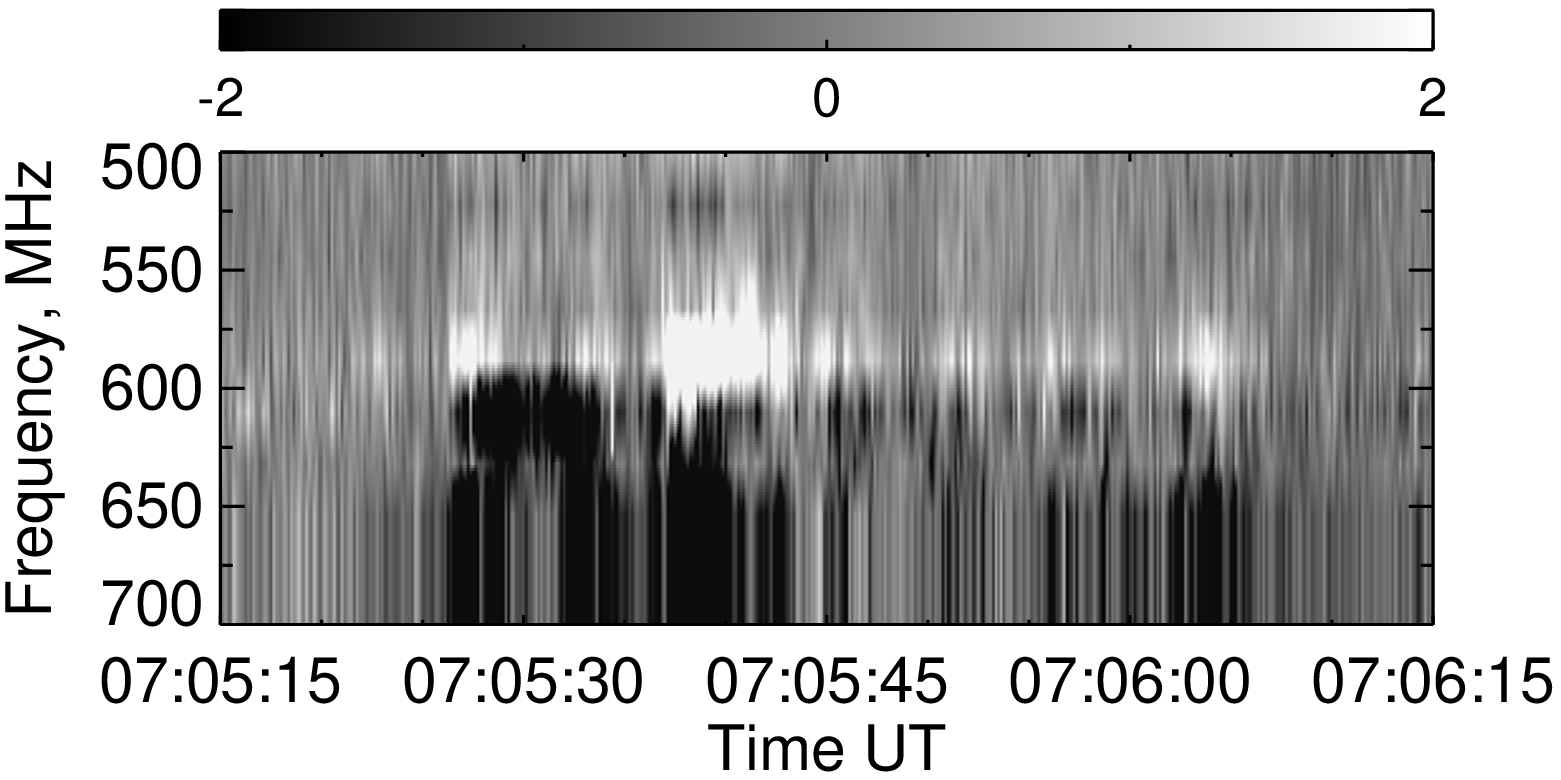}
  }
 \caption{The {MUSER} dynamic spectra of the intensity $R+L$ (left) and polarisation $R-L$ (right). Colour bars indicate the corresponding fluxes in s.f.u. The horizontal line in the left panel indicates the frequency where the dynamic spectrum reaches its maximum value.}
 \label{fig:DynSpectrum}
\end{figure}

Figure~\ref{fig:DynSpectrum} presents the dynamic spectrum by MUSER both for intensity ($R+L$) and Stokes parameter $V$ or circular polarisation ($R-L$). The intensity spectrum on the left panel shows a series of {narrow-band}, quasi-periodic bursts similar to the type I bursts, which are known to have narrow bandwidth and short duration  \cite[see, for example,][]{Storms1985srph.book..415K}. 
Classical type I noise storms are usually observed below 400--450~MHz, {corresponding to coronal altitudes around 0.2 solar radii} \citep{typeI_Nancey2015A&A...576A.136M}. We note that type I radio bursts are associated with non-thermal electron emission. This fact allows us to assume a similar emission mechanism for the considered decimetric bursts.

The decimeter bursts are polarised, as one can see in the right panel of Figure~\ref{fig:DynSpectrum}  presenting the Stokes $V$ parameter. 
The presence of polarisation indicates the non-thermal origin of the emission. The sign of polarisation does not change with time. It is mostly positive, but we can see the sign inverts at frequencies below $\approx$ 600~MHz.
This inversion of the polarisation sign indicates that the emission at those frequencies could originate near the top of a high loop (substantially higher the one observed in MW in Figure~\ref{fig:EUV}).

\section{Quasi-periodic properties of the microwave and decimeter emission}
\label{s:qpp}
The microwave emission (Figure~\ref{fig:GOES}) shows quasi-periodic behaviour with the period around 20--30~s during the flux rise phase from 07:04:10~UT to 07:06:30~UT. The dynamic spectra shown in Figure~\ref{fig:DynSpectrum} also demonstrate quasi-periodic enhancements of the decimeter emission which is clearly seen in the intensity from 07:05:22~UT to 07:06:05~UT, \textit{i.e.} around maximum of the microwave flux (at 07:05:40~UT), with a shorter period.
In this section, we perform a detailed analysis of the periodic properties of the MW and decimeter intensities shown in Figures~\ref{fig:GOES} and \ref{fig:DynSpectrum}.
The technique which we apply for the detection of QPPs in the MW and decimetric time profiles consists of removing their low-frequency trends, and processing the detrended time series using decomposition techniques. 

The problem of detrending in the context of QPP detection, especially in the presence of coloured noise, is described in, for example, \citet{2018ApJ...858L...3K}. Different techniques currently used for the detection and analysis of QPP signals in solar flares were summarised and comprehensively tested in \citet{2019ApJS..244...44B}, see also \citet{2020STP.....6a...3K}. In our study,
we determine the low-frequency trends of the observed emissions by
smoothing the original time profiles through the convolution with a polynomial function (also known as a Savitzky–Golay filter),
applied with a smoothing time window of the width $\tau$.
{We show this width in Figure~\ref{fig:Wavelet_BBMS} and Figure~\ref{fig:Wavelet_MUSER} by the horizontal dotted lines.}
After subtracting the trend, the detrended time series were analysed with both the wavelet (Morlet) transform \citep{1998BAMS...79...61T} and the Fourier periodogram. The statistical significance level in the wavelet analysis was obtained for the red noise function. The significance level for the Fourier periodogram is calculated according to \citet{1986ApJ...302..757H}.
Such a combination of the Fourier and wavelet approaches accompanied by a rigorous assessment of the statistical significance of the obtained periodic components in comparison with the background noise allows us to exclude possible artifacts of the data processing. 

\subsection{QPPs at microwaves}
\label{s:qpp_MW}

\begin{figure} 
 \centerline{
 \includegraphics[width=0.8\textwidth,clip=]{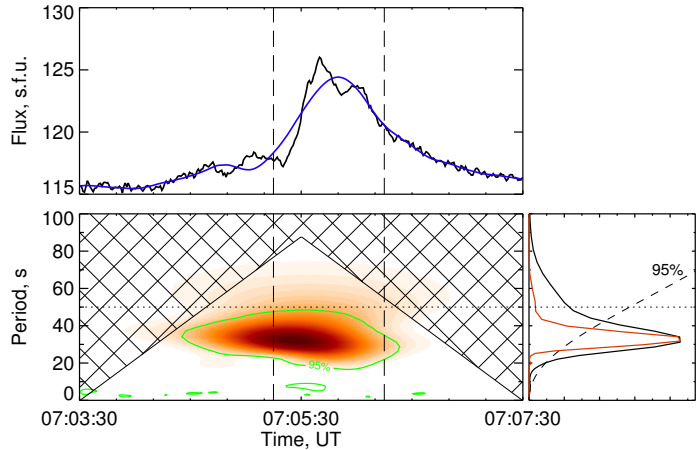}
  }
 \caption{Top panel: the BBMS flux at 8.056~GHz (black curve) and its low-frequency trend (blue curve). Bottom left: the Morlet wavelet power spectrum (gradient of red) of the de-trended flux component. Bottom right: the corresponding global wavelet (black line) and Fourier (red line) spectra. The significance level is shown by the green contour and the dashed line in the bottom left and right panels, correspondingly. The horizontal lines indicate the width $\tau=50$\,s of the smoothing window used for obtaining the trend. The vertical dashed lines show the time interval {07:05:15--07:06:15~UT where the  decimetric QPPs were found (see Figure~\ref{fig:Wavelet_MUSER}).}}
 \label{fig:Wavelet_BBMS}
\end{figure}

An example of the original time profile of the BBMS flux at 8.056~GHz, its low-frequency trend obtained for $\tau = 50$~s, and the corresponding wavelet and Fourier power spectra calculated after removing the trend are shown in Figure~\ref{fig:Wavelet_BBMS}.
As such, the wavelet analysis detects a single-mode QPP signal with the characteristic period around $30 \pm 10$~s. 
The Fourier periodogram of the auto-correlation function of the detrended signal reveals the same period $30 \pm 5$~s. 

\begin{figure} 
 \centerline{
 \includegraphics[width=0.5\textwidth,clip=]{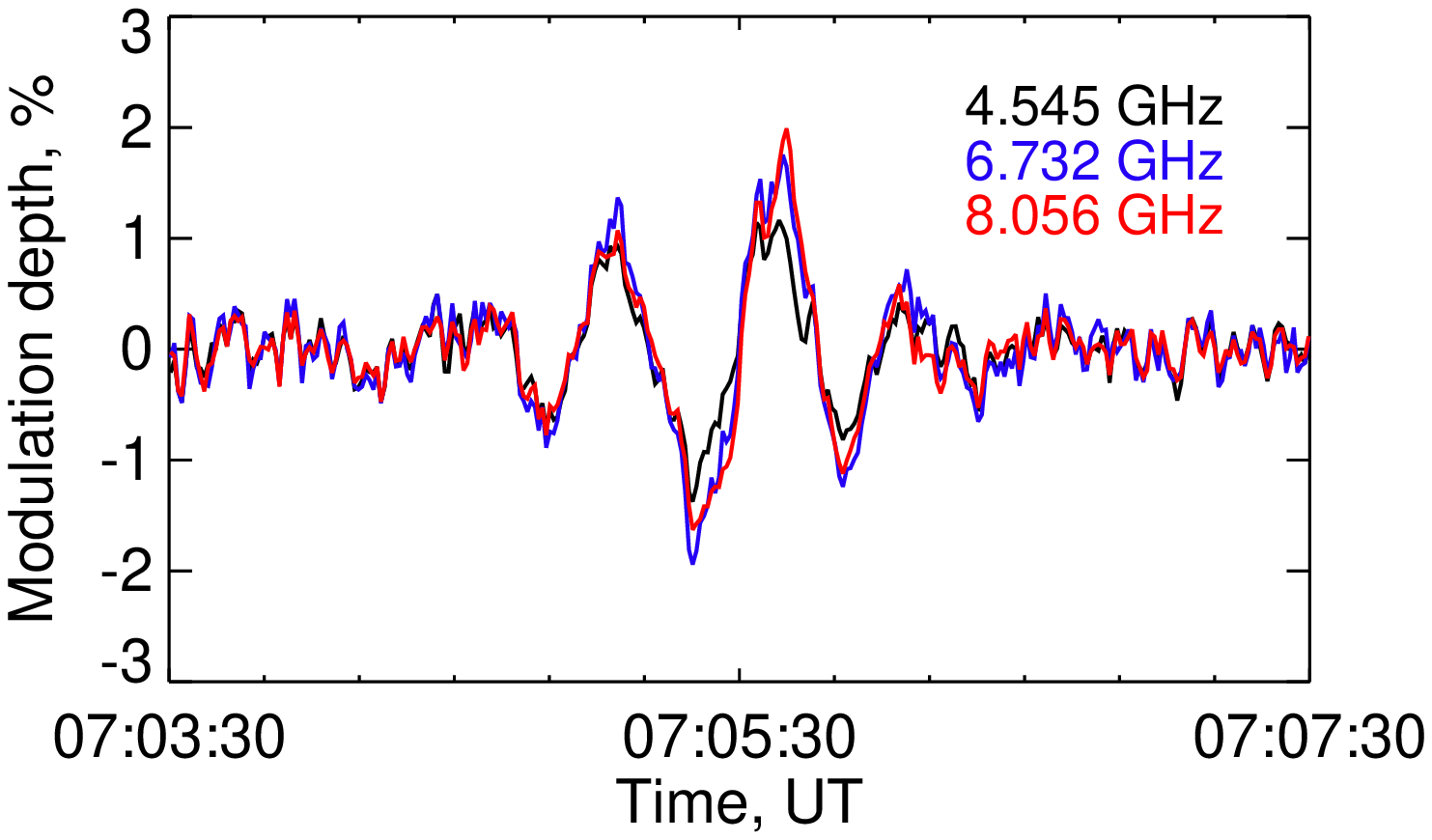}
  \includegraphics[width=0.5\textwidth,clip=]{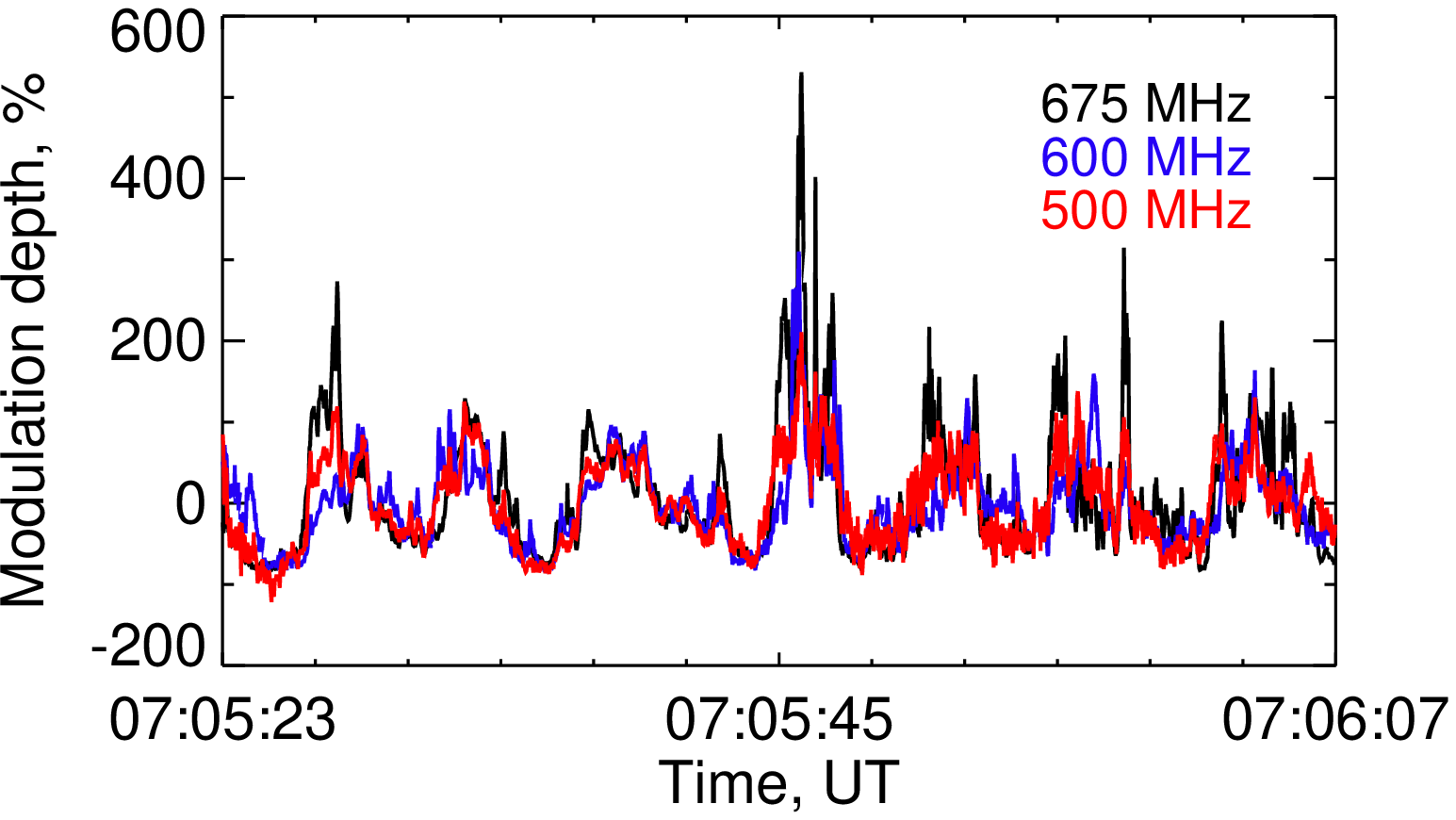}
  }
 \caption{Comparison of the de-trended oscillation time profiles at different frequencies: modulation depths of the fluxes in the microwave (left panel) and decimeter (right panel) bands.
 }
 \label{fig:Detrended}
\end{figure}

These quasi-periodic variations with the same period about 30\,s are detected in fluxes at all microwave frequencies within the 4--8~GHz band. Moreover, the QPPs for all these time profiles demonstrate an in-phase behaviour, as shown by Figure~\ref{fig:Detrended}. Here, the modulation depth, which is defined as the ratio of the detrended flare light curve to its low-frequency trend, is illustrated for three frequencies; the optically thick part of the MW spectrum (4.545~GHz, see the left panel in Figure~\ref{fig:EUV}), near the spectral maximum (6.732~GHz), and in the optically thin part of the spectrum (8.056~GHz). The modulation depths are found to be around a few percent for all the profiles. The local differences are probably caused by noise in the raw data and time binning.

\subsection{QPPs at decimeters}
\label{s:qpp_dm}
\begin{figure} 
 \centerline{
 \includegraphics[width=0.8\textwidth,clip=]{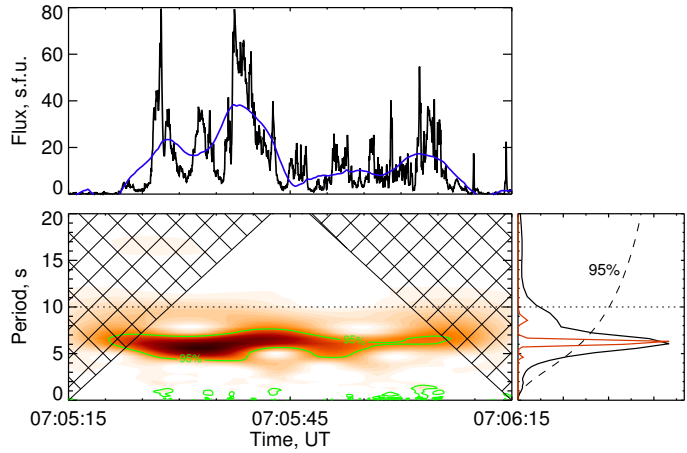}
  }
 \caption{The same as in Figure~\ref{fig:Wavelet_BBMS} but for the flux registered by MUSER at 675~MHz.}
 \label{fig:Wavelet_MUSER}
\end{figure}

A QPP signal with a shorter period of $6\pm 1.5$~s and at least seven well-pronounced oscillation cycles is found in the time profiles obtained with MUSER at all frequencies from 500~MHz to 700~MHz. This period is well seen in the original time profiles, wavelet spectra, and Fourier periodograms, as demonstrated in Figure~\ref{fig:Wavelet_MUSER} for the radio flux intensity at 675~MHz, where the amplitude of the flux and its variations reach their maxim{a} (see Figure~\ref{fig:DynSpectrum}). The results are the same for different smoothing widths $5$~s~$ < \tau < 20$~s, while the flux variations with longer timescales are associated with the relaxation of the local plasma conditions after a primary energy release. For all radio frequencies, the 6~s QPPs are dominant and statistically significant in the wavelet spectrum. Similarly to the QPPs in microwaves, all the QPPs in the decimeter range are in-phase, with a modulation depth around 100\% (see Figure~\ref{fig:DynSpectrum} and Figure~\ref{fig:Detrended}).






\section{Discussion and conclusion}
\label{s:DiscussionC}

We performed analysis of quasi-periodic pulsations (QPP) simultaneously observed in the microwave and decimetric emissions of SOL2017-09-05T07:04 C6.9 solar flare to explore and establish the relationship between these two radio emission mechanisms. More specifically, we found the following properties of the QPP events in microwaves:

\begin{itemize}
    \item Three clear cycles with an oscillation period of about 30~s.
    \item Both the peak frequency of the microwave spectrum and photon spectral index did not change during the QPPs.
    \item The spectral index at low frequencies indicates the absence of the Razin effect.
    \item Time profiles of the MW oscillations demonstrate an in-phase behaviour both in the optically thick and optically thin emissions.
    \item The modulation depth of the oscillations is similar at different frequencies, and its values are about a few percent.
\end{itemize}

and in decimeters:
\begin{itemize}
    \item At least seven oscillation cycles with an oscillation period of about 6~s.
    \item The frequency of the maximum intensity is about 675~MHz, and this did not change during the event.
   \item The QPPs appear in the narrow spectral band 500--700~MHz.
    \item The sign of the Stokes $V$ parameter did not change in time during the event, but it reversed at the lowest frequency.
    \item Time profiles of the decimeter oscillations demonstrate an in-phase behaviour.
    \item The modulation depth of the oscillations is around 100\%.
\end{itemize}


\begin{figure} 
 \centerline{
 \includegraphics[width=\textwidth]{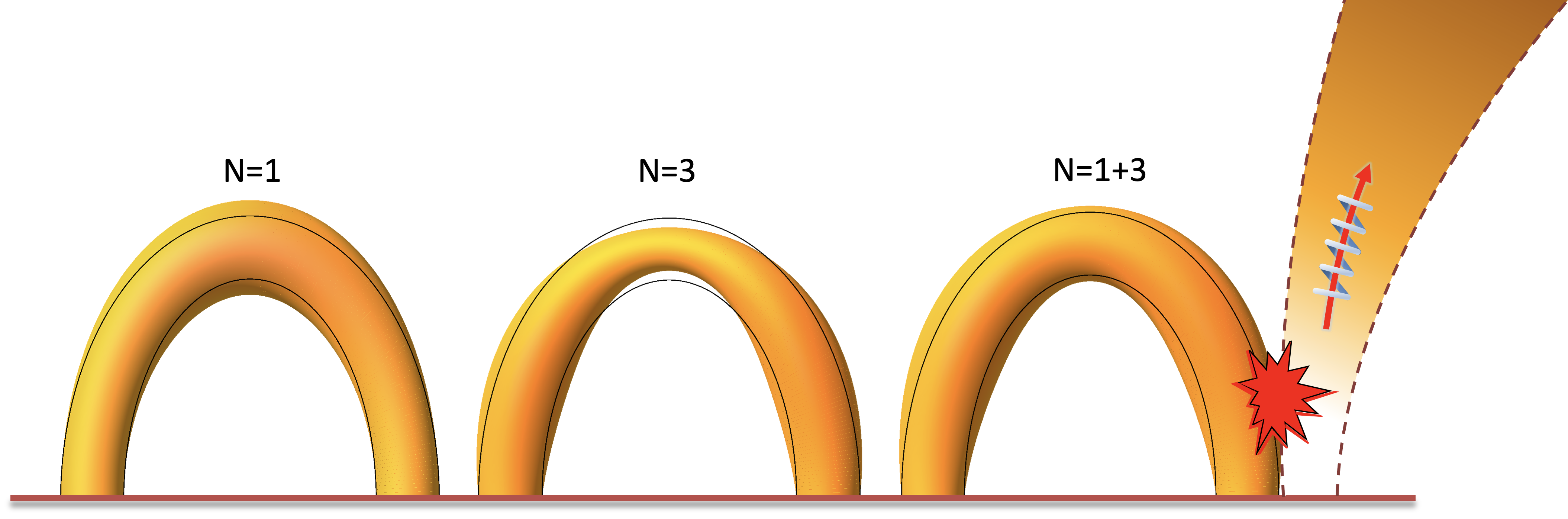}
}
 \caption{A schematic illustration of a spatial structure of the loop boundary perturbed by a fundamental harmonic ($N=1$), third harmonic ($N=3$), and their superposition ($N=1+3$) of a standing fast sausage wave. The perturbations near the loop legs, caused by the third sausage harmonic, allow for the periodic interaction with a nearby plasma structure and acceleration of charged particles. {In our scenario, the external plasma structure is considered to be a closed magnetic loop, substantially higher than the sausage-oscillating MW loop.}
 The thin black lines show the unperturbed state of the loop.
 The choice of the colour scheme is arbitrary.}
 \label{fig:sketch}
\end{figure}

The observed oscillation period $P_1=30$\,s and an in-phase behaviour of the QPP signals in the MW band are in good agreement with the modulation of the gyrosynchrotron emission by a sausage mode of a standing fast magnetoacoustic wave in the coronal loop \citep[see \textit{e.g.}][for observational and modelling examples of this scenario]{2005A&A...439..727M, 2008A&A...487.1147I, Kuznetsov2015SoPh}.
{Moreover, impulsive energy releases could naturally excite several harmonics of the sausage mode simultaneously in the coronal loop.}
The first and the most powerful harmonic of the sausage mode is known to have nodes near the loop legs and anti-nodes (\textit{i.e.} maximum perturbation amplitude) at the top of the flaring loop (see Figure~\ref{fig:sketch}, for a schematic illustration). Likewise, the MW emission at the frequencies around the MW spectral peak and below are also known to be stronger near the loop top \citep[see \textit{e.g.} Figure~1 in][]{1998ARA&A..36..131B}. Hence, the fundamental sausage harmonic should dominate at these MW frequencies, as it is observed at 4--8\,GHz in this work. No signatures of higher sausage harmonics were detected in the MW band.


As we demonstrated in Section~\ref{s:Observations}, the flare loop seen in MW is far too low to emit in the decimetre range.  Hence, we are not able to use a single-loop model for the interpretation of the 6 s QPP signal at 500--700\,MHz. On the other hand, the MW flare loop is seen in Figure~\ref{fig:EUV} to situate nearby an apparently open or high loop structure seen in the 174\,\AA\ EUV channel. More specifically, the lower part of the MW loop, observed at 6.8 and 7.5\,GHz, is seen to be closer to this external loop-like structure than the loop top. We note here that the MW sources at 6.8 and 7.5\,GHz indicate not the footpoint of the flare loop but a segment above it. Thus, we suggest that the observed 6 s decimetric QPPs could be attributed to the effect of a higher sausage harmonic of the flare loop, which has non-zero perturbation amplitudes near the loop legs, thus allowing for an effective periodic interaction of the sausage-oscillating loop with this external plasma structure and formation of magnetic X-points and current sheets, as schematically illustrated in Figure~\ref{fig:sketch}. 

For example, the third harmonic of a sausage-oscillating flare loop is expected to have a period of $P_3=P_1/3\approx 10$\,s {\citep[or $\gtrsim 10$\,s taking the effect of waveguiding dispersion into account, see \textit{e.g.}][for details]{2009A&A...503..569I,2016RAA....16...92Y}}. Following a qualitative scenario developed by \citet{2006A&A...452..343N}, this essentially transverse very small-amplitude wave modulates the plasma conditions at the interface between the sausage-oscillating loop and the external loop, causing, in particular, strong variations of the local electric current density with the period of the mother higher-harmonic sausage oscillation, \textit{i.e.} around 10~s. For the local current density periodically exceeding some threshold value, the plasma is known to be subject to various micro-instabilities and micro-turbulence, which can in turn lead to anomalous values of the plasma resistivity and thus provide plausible conditions for periodic triggering of magnetic reconnection, acceleration of charged particles and hence bursts of low-frequency coherent radio emission. Moreover, as the extreme values of the current density are achieved at both positive and negative half-cycles of the oscillation \citep[see Figure~3 in][]{2006A&A...452..343N}, the rate at which magnetic reconnection triggers in this scenario is two times higher than the 10 s period of the higher-harmonic sausage wave inducing it. In other words, the third harmonic of the sausage-oscillating loop with a 10 s period is expected to trigger magnetic reconnection that can cause the acceleration of charged particles every 5~s in this scenario, which is very close (or identical within the uncertainties that we have in the data analysis) to the observed 6 s period in the decimetric band.

It is also worth mentioning here that in addition to perturbing the loop near the loop legs where the interaction with the external plasma structure potentially takes place, the third sausage harmonic has a non-zero perturbation amplitude near the loop top where the maximum of the MW emission at 4--8\,GHz is observed. However, we detected no signatures of this higher harmonic in the MW observations, mainly due to its low amplitude in comparison with the fundamental sausage harmonic dominating in this part of the loop.
{In any case, the suggested interpretation should be considered as a hypothesis only, as it is very difficult to find the direct evidence of reconnection in observations. On the one hand, it is caused by the lack of the EUV data available for this event. On the other hand, previous studies demonstrated that small-scale quasi-periodic reconnection does not necessary change the global magnetic topology of the flare region \citep[see e.g.][]{2000A&A...360..715K, 2009A&A...494..329M}.}
{We also note that the propagating fast magnetoacoustic waves cannot be completely ruled out in this interpretation. For example, the detected form of the wavelet power spectrum shown in Figure~\ref{fig:Wavelet_BBMS} resembles a characteristic tadpole structure typical for quasi-periodic fast-propagating wave trains \citep[see e.g.][]{2004MNRAS.349..705N, 2014ApJ...788...44M, 2021MNRAS.505.3505K}. On the other hand, it is less clear how to explain the observed ratio of the oscillation periods in MW and decimeters by propagating waves.}


{The oscillation modulation depth found in the microwave emission does not exceed 2--3$\%$, while the oscillation modulation depth detected in the decimeter band is about 100$\%$. The modulation depth found in the microwave emission agrees with the results obtained by modelling of the microwave emission from a sausage-oscillating loop, which showed the modulation depth around  1--5$\%$ \citep{Kuznetsov2015SoPh}.  The observed disparity in the microwave and decimeter modulation depths is likely to occur due to the difference in the emission generation mechanisms.} 
{Namely, the microwave emission is dependent upon many parameters such as accelerated electron flux,  magnetic field, background plasma density, source size and flare loop inclination \citep[see][for details]{Dulk1985ARA}, which for some combinations could lead to the observed relatively low modulation amplitudes.  For the coherent plasma emission, this is likely stimulated by deca-keV electrons in a much larger loop structure.  Each burst of radio emission is potentially caused by a separate acceleration event, creating the high modulation depth of 100\%.}

{To the best of knowledge, this work presents the first collaborative analysis of the data from SRH and MUSER. It allowed us to establish the relationship between the oscillations detected in the MW and decimetric emission of the solar flare.
The scenario explaining the obtained periodicities is similar to that considered by \cite{Benz2011SoPh..273..363B}, with the magnetic topology of the event assuming the current sheet that generates the decimetric emission is located not above the loop-top source but nearby the flare loop. Clearly, the proposed physical model is rather qualitative than quantitative and needs further validation by using comprehensive theoretical simulations {and forward modelling}, addressing, in particular, a huge disparity in the observed QPP modulation depths in the MW and decimetric bands.
Despite this, the work demonstrates a clear potential of multi-band radio observations for deeper understanding of the mechanisms of particle acceleration and propagation during solar flares and the role of MHD waves in these processes. }




%

%

%

%
 \begin{acks}
This study was funded by the RFBR (project number 21-52-10012) and the Royal Society (project number IEC\textbackslash R2\textbackslash 202175). The work of LK, AK and DK was partly  financially supported by the Ministry of Science and Higher Education of the Russian Federation.
The work of EK and DK on the detection of quasi-periodic signals and interpretation of results was supported by Russian Scientific Foundation grant 21-12-00195. DK was also supported by the STFC consolidated grant ST/T000252/1. CT was was supported by the NSFC grants 11790301 {and 11941003. Authors thank CSUC ’Angara’ for BBMS data, teams operating SRH-48, MUSER, PROBA2/SWAP, RSTN and GOES for open access to observational data.}
 \end{acks}

%
%
\bibliographystyle{spr-mp-sola}
\bibliography{ref_nonst_qpp}  
%
%
%
%

\end{article} 
\end{document}